\documentclass[journal=aesccq,manuscript=article, layout=twocolumn]{achemso}

\usepackage{chemformula}
\usepackage[T1]{fontenc}

\SectionNumbersOn

\usepackage{subcaption}

\let\oldmaketitle\maketitle
\let\maketitle\relax

\author{Marten T. Raaphorst}
\email{m.t.raaphorst@lic.leidenuniv.nl}
\author{Joan Enrique-Romero}
\author{Thanja Lamberts}
\email{a.l.m.lamberts@lic.leidenuniv.nl}
\affiliation[LIC]
{Leiden Institute of Chemistry, Gorlaeus Laboratories, Leiden University, PO Box 9502, 2300 RA Leiden, The Netherlands}
\alsoaffiliation[Obs]
{Leiden Observatory, Leiden University, P.O. Box 9513, 2300 RA Leiden, The Netherlands}
\title[]
  {Toward Unraveling Cyanopolyyne Surface Chemistry: A Preview on Isolated Systems From \ch{HC3N} to Ethyl Cyanide and Propylamine}


\begin{document}

\twocolumn[
    \begin{@twocolumnfalse}
    \oldmaketitle
    \begin{abstract}
    Cyanopolyynes, a family of nitrogen containing carbon chains, are common in the interstellar medium and possibly form the backbone of species relevant to prebiotic chemistry. 
    Following their gas phase formation, they are expected to freeze out on ice grains in cold interstellar regions.
    In this work we present the hydrogenation reaction network of isolated \ch{HC3N}, the smallest cyanopolyyne, that consists over-a-barrier radical-neutral reactions and barrierless radical-radical reactions. 
    We employ density functional theory, coupled cluster and multiconfigurational methods to obtain activation and reaction energies for the hydrogenation network of \ch{HC3N}.
    This work explores the reaction network of the isolated molecule and constitutes a preview on the reactions occurring on the ice grain surface.
    We find that the reactions where the hydrogen atom adds to the carbon chain at carbon atom opposite of the cyano-group give the lowest and most narrow barriers. 
    Subsequent hydrogenation leads to the astrochemically relevant vinyl cyanide and ethyl cyanide. 
    Alternatively, the cyano-group can hydrogenate via radical-radical reactions, leading to the fully saturated propylamine. 
    These results can be extrapolated to give insight into the general reactivity of carbon chains on interstellar ices.
    
    \end{abstract}

    \vspace{1cm}
    
    \end{@twocolumnfalse}
]

\section{Introduction}

Interstellar space makes a harsh environment for molecules to meet and chemical reactions to occur due to the extremely low densities and temperatures. Despite this, the interstellar medium (ISM), contains a rich chemistry\cite{herbst_complex_2009, caselli_our_2012, mcguire_2021_2022}. In cold regions (temperatures as low as $\sim$10 K), such as dense clouds and prestellar cores, molecules like \ch{H2O}, \ch{NH3}, \ch{CO2}, and \ch{CH3OH} form on the surfaces of dust grains, covering the dust particles with an icy molecular mantle \cite{leger_does_1983, benson_survey_1989, hudgins_mid_1993, crapsi_observing_2007, linnartz_solid_2011, vandishoeck_astrochemistry_2014, boogert_observations_2015, mcclure_ice_2023}. 

On the other hand, unsaturated carbon chains, molecules consisting of mainly carbon in a linear chain, form predominantly in the gas phase\cite{herbst_complex_2009, herbst_synthesis_2017, ceccarelli_organic_2023, taniguchi_carbonchain_2024}. A particular class of carbon chains are the cyanopolyynes, with the general molecular formula H$-$(C$\equiv$C)$_{n}-$C$\equiv$N (and \textit{n} = 1,2,3...). Cyanopolyynes up to length of 11 carbon atoms have been observed in different astronomical regions, most notably at the cyanopolyyne peak position in the molecular cloud TMC-1 \cite{turner_detection_1971, avery_detection_1976, morris_cyanoacetylene_1976, zuckerman_crl_1976, little_detection_1977, winnewisser_detection_1978, broten_detection_1978, kroto_detection_1978, kunde_c4h2_1981, mauersberger_dense_1990, bockelee-morvan_new_2000, chapillon_chemistry_2012}. In such a dense cloud, any species heavier than helium, for instance CO, but also carbon chains and cyanopolyynes, eventually freeze out on the surfaces of ice-covered grains\cite{jaberal-edhari_history_2017, taniguchi_cyanopolyyne_2019, caselli_central_2022}. The subsequent solid-state chemistry of the adsorbed cyanopolyynes has been hypothesized to link to the formation of prebiotic molecules. Those species typically consist of a carbon backbone and include nitrogen (and oxygen) atoms, such as glycine (\ch{C2H5NO2}) and adenine (\ch{C5H5N5}) \cite{balucani_elementary_2009}, but more specifically, cyanopolyynes have been suggested to be precursors to fatty acids and cyclic species \cite{oro_chemical_1995a, shingledecker_detection_2021}. However, the reactivity of carbon chains and cyanopolyynes on ice grains has thus far not been explored \cite{taniguchi_carbonchain_2024, jaberal-edhari_history_2017, taniguchi_cyanopolyyne_2019}. We hypothesize that hydrogenation of cyanopolyynes on ice grains leads to precursors of prebiotic molecules. 

The smallest cyanopolyyne is cyanoacetylene (\ch{HC3N}) and can be used as a case study on solid-state carbon chain reactivity. \ch{HC3N} was first detected in 1971 in the Sgr B2 molecular cloud and has since then been observed in many interstellar environments \cite{turner_detection_1971, morris_cyanoacetylene_1976, bujarrabal_abundance_1981, chapillon_chemistry_2012, thiel_complex_2017, bianchi_cyanopolyyne_2023}. It is commonly found in dark molecular clouds \cite{morris_cyanoacetylene_1976, snell_observations_1981, winstanley_cyanopolyyne_1996a, miettinen_malt90_2014, bianchi_cyanopolyyne_2023} and upon adsorption on icy grains, cyanoacetylene can in principle react with hydrogen atoms. Such hydrogenation reactions on ice surfaces are common and have been studied in detail for other molecules.\cite{herbst_complex_2009, kobayashi_hydrogenation_2017, zaverkin_tunnelling_2018, molpeceres_radical_2022} For instance, water can be formed by successive reactions of O atoms with H atoms or \ch{H2} molecules \cite{cuppen_simulation_2007, dulieu_experimental_2010}. 
Successive hydrogenation of \ch{HC3N} potentially leads to vinyl cyanide (\ch{CH2CHCN}) and ethyl cyanide (\ch{CH3CH2CN}), both observed in the ISM in the gas phase \cite{gardner_detection_1975, johnson_detection_1977}. Whether this hydrogenation actually occurs on the ice grains has been a point of discussion for some time \cite{blake_molecular_1987, minh_upper_1991, dickens_hydrogenation_1997, irvine_extraterrestrial_1998, belloche_increased_2009, calcutt_almapils_2018}.
A recent study based on ALMA observations, however, points to vinyl and ethyl cyanide originating from prestellar ices\cite{nazari_nbearing_2022}. Additionally, ethyl cyanide has been tentatively observed in interstellar ice with JWST \cite{nazari_hunting_2024}. The fully saturated species, propylamine (\ch{CH3CH2CH2NH2}) has as of yet not been observed in the ISM.

Although specific reactions and intermediates in the hydrogenation reaction network starting with \ch{HC3N} have already been studied \cite{woon_rate_1997, fukuzawa_molecular_1997, huang_crossed_2000, balucani_crossed_2000, parker_kinetics_2004, singh_quantum_2021, arnau_initio_1993}, the energetics of the full network have not been reported. Therefore, chemical pathways are overlooked and accurate activation energies are not available for astrochemical models \cite{belloche_increased_2009}. 

In this work we present a computational study on the hydrogenation network of cyanoacetylene, starting from \ch{HC3N} and reaching vinyl cyanide (\ch{CH2CHCN}), ethyl cyanide (\ch{CH3CH2CN}) and propylamine (\ch{CH3CH2CH2NH2}).
Our aim is to (a) understand the reactivity of \ch{HC3N} specifically, and (b) extrapolate it to predict hydrogenation of carbon chains and cyanopolyynes on interstellar ices in general to (c) provide insight into the formation of complex organic molecules and possible precursors for the molecules of life. 
We studied the energetics of the network, specifically the activation and reaction energies, by employing computational chemical techniques. Based on barrier height, barrier width and crossover temperatures, we report which reactions are likely to occur under cold ISM conditions. This work provides a benchmark and first insights into the complicated solid-state chemical network of cyanoacetylene hydrogenation. We consider the isolated reactions, i.e., only the H$_n$C$_3$N species with an H atom, as a preparatory work to include surface molecules or surface effects in a forthcoming publication. The work is structured as follows: section \ref{sec: methods} describes the computational methods and details of the calculations. Section \ref{sec: results} starts with results obtained for the first hydrogenation step of \ch{HC3N}. This is followed by the description of the first radical-radical reaction \ch{H2C3N + H}. Then we present the pathways towards ethyl cyanide via vinyl cyanide and the multiple possible pathways towards propylamine formation. The section ends with some astrochemical implications. Section \ref{sec: conclusions} concludes with a summary and outlook for future work.

\section{Methods} \label{sec: methods}

All density functional theory (DFT) calculations were carried out with ORCA 5.0.4 \cite{neese_orca_2012, neese_orca_2020, neese_software_2022}. We chose what functional to use in this work after running a benchmark study in which we used both Molpro 2022.3\cite{werner_molpro_2012, werner_molpro_2020, MOLPRO} and OPENMOLCAS 21.02\cite{fdez_galvan_openmolcas_2019, li_openmolcas_2023} to run CCSD(T)-F12 \cite{alder_new_2007, knizia_simplified_2009} and CASPT2 calculations \cite{vancoillie_parallelization_2013}, respectively. DFT calculations included geometry optimizations, transition state searches, intrinsic reaction coordinates (IRC) and zero-point vibrational energy calculations. All stationary points were checked as such by frequency calculations. Unless otherwise specified, we used ORCA's TightSCF setting and default DefGrid2 integration grid. All reported single point energies were obtained at the uCCSD(T)-F12/cc-pVTZ-F12 level \cite{peterson_systematically_2008}. Additionally, all activation and reaction energies are given relative to the asymptotic state, i.e. the molecule and H atom(s) infinitely separated. Two different types of reactions appear in this study: barrier-mediated and barrierless ones in open-shell singlet systems. In the following, we provide further details.

\subsection{Barrier-mediated reactions}

For the reactions taking place over a barrier, we optimized the geometries with the MPWB1K density functional \cite{zhao_hybrid_2004}, made available to ORCA with use of LibXC \cite{lehtola_recent_2018}, in combination with the D4 dispersion correction \cite{caldeweyher_extension_2017, caldeweyher_generally_2019} and the def2-TZVP basis set \cite{weigend_balanced_2005}. 

\begin{figure}
    \centering
    \includegraphics[width=240pt]{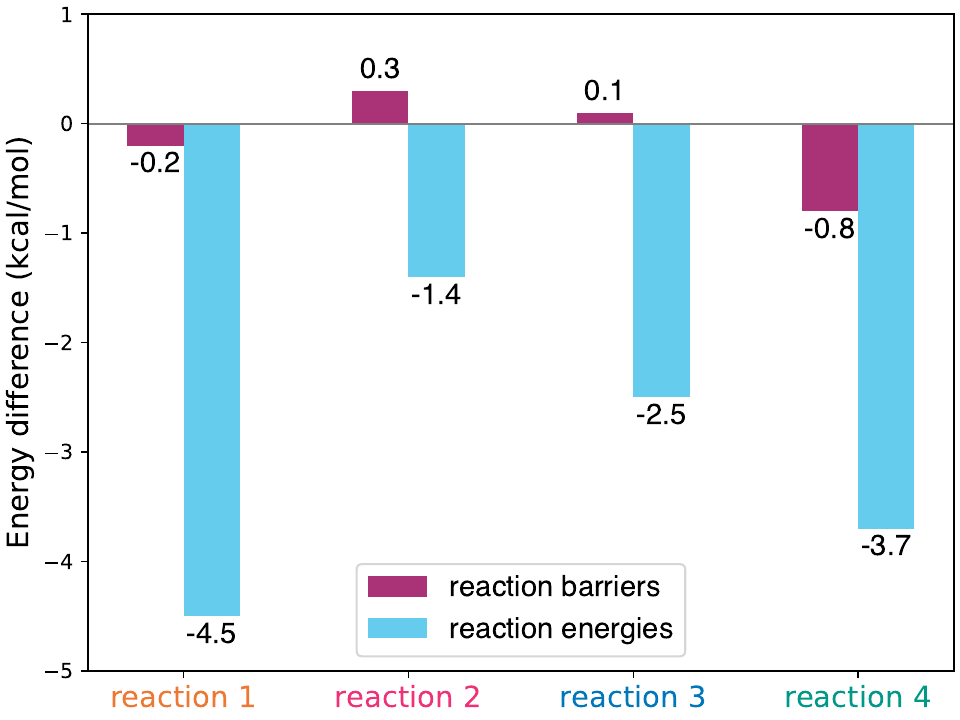}
    \caption{Energy differences between the DFT MPWB1K-D4/def2-TZVP values and uCCSD(T)-F12/cc-pVTZ-F12 values. ZPE corrections are not included. The geometries were optimized with MPWB1K-D4/def2-TZVP. The labels reaction 1-4 correspond to H addition to the C1, C2 and C3 carbons, and addition to the nitrogen atom respectively (see also Figure \ref{fig: HC3N+H energy diagram}).}
    \label{fig: HC3N+H MPWB1K benchmark}
\end{figure}

We performed a benchmark study on the MPWB1K functional for the four addition reactions \ch{HC3N} + H. We calculated the reaction energies and heights of the reaction barriers both at the MPWB1K-D4/def2-TZVP and uCCSD(T)-F12/cc-pVTZ-F12 level, see Figure \ref{fig: HC3N+H MPWB1K benchmark}. The MPWB1K functional in combination with the D4 dispersion correction accurately captures the height of the reaction barriers, which are our main interest given the low-temperature conditions the reactions take place in. Additionally, we compared the energies obtained with 6 other density functionals to the coupled cluster results, see Figure S1. We found that the MPWB1K functional gives the most accurate barrier heights.

\subsection{Barrierless reactions}

We studied the radical-radical reactions with broken symmetry DFT, which is a method of approximating multi-reference character within DFT \cite{noodleman_valence_1981, neese_definition_2004}. This method has previously been shown to be effective for describing these types of reactions \cite{enrique-romero_revisiting_2020}.
To investigate the different possible products of a radical-radical reaction, we performed a series of geometry optimizations (ranging from 93 to 114) per reaction. For each optimization the initial geometry consisted of a hydrogen atom placed somewhere around the H$_n$C$_3$N radical, at a distance of 3.0, 3.5 or 4.0 \AA\ away from the nearest atom of the radical. During the geometry optimization the system converges to one of the possible closed-shell products. The entire series of optimizations then indicates the possible product channels. If performed on an ice surface, these calculations could also indicate whether the H additions are more efficient for certain incoming directions. However, as a result of our calculations being performed in vacuum, we refrain from publishing quantitative branching ratios. This will be the topic of a dedicated follow-up work on interstellar ice analogues.

\begin{figure*}[h]
    \centering
    \begin{subfigure}{0.48\textwidth}
            \includegraphics[width=220pt]{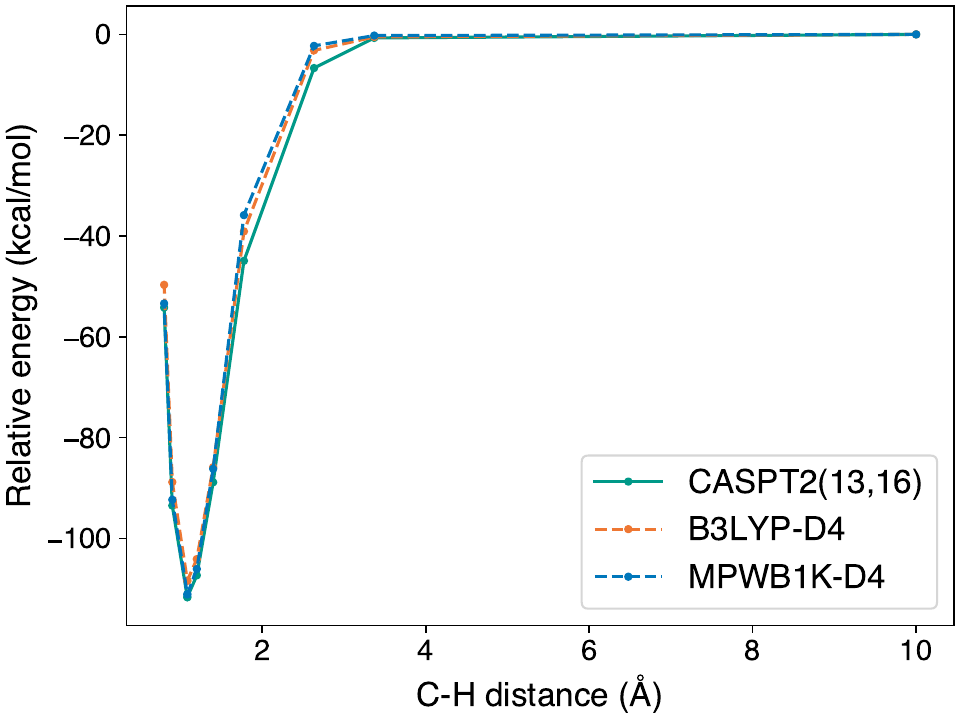}
    \caption{}\label{fig: H2C3N+H C-side}
    \end{subfigure}
        \begin{subfigure}{0.48\textwidth}
            \includegraphics[width=220pt]{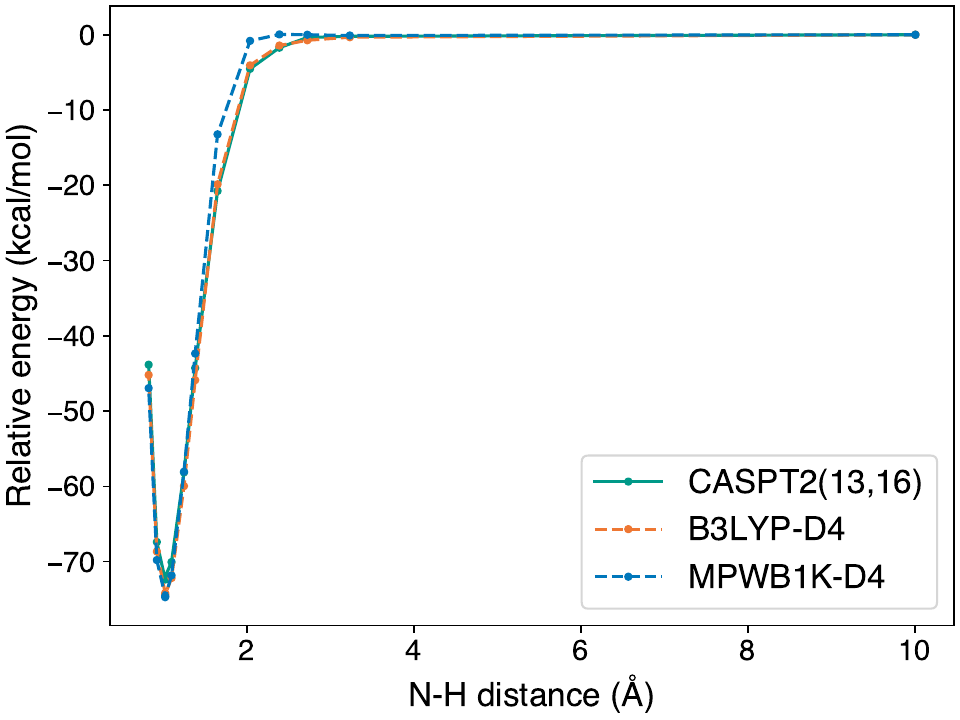}
    \caption{}\label{fig: H2C3N+H N-side}
    \end{subfigure}
    \caption{Energy plots for scans of the reaction \ch{H2C3N}+H, obtained with three different methods: B3LYP-D4/def2-TZVP, MP1B1K-D4/def2-TZVP and CASPT2/cc-pVTZ. The left (\ref{fig: H2C3N+H C-side}) plot shows the H addition to the C2 carbon and the right (\ref{fig: H2C3N+H N-side}) shows the H addition to the N atom. The energies are given relative to the state at C-H or N-H distance of 10 \AA\ and were obtained with single point calculations on geometries optimized at the B3LYP-D4/def2-TZVP level.}
    \label{fig: H2C3N+H benchmark}
\end{figure*}

The broken-symmetry calculations were performed at the B3LYP-D4/def2-TZVP level, benchmarked to correctly describe the energy trend of the radical-radical reactions compared to CASPT2 calculations. The CASPT2 calculations were performed using the cc-pVTZ basis set \cite{dunning_gaussian_1989} and an active space including 16 electrons and 13 orbitals. We chose the orbitals such that: 1) we were as close to a full valence active space as possible 2) all the orbitals partaking in the reaction are included 3) the active space remained the same for the geometries along the benchmark scan, in other words, making sure the orbitals formed in the reaction are included in the active space. The benchmark (see Figure \ref{fig: H2C3N+H benchmark}) shows that B3LYP-D4 more closely captures the potential energy surface for the \ch{H2C3N} + H reaction compared to MPWB1K-D4, in particular regarding the lack of a barrier along the N-H bond formation. We emphasize that it is specifically the barrierless-character of the PES that makes B3LYP-D4 favorable over MPWB1K-D4, as an erroneous barrier would greatly influence the branching ratio of the product channels described in the next sections. For this reason we opt to use MPWB1K-D4 for all barrier-mediated reactions and B3LYP-D4 for all barrierless reactions.

\section{Results and discussion}\label{sec: results}

\subsection{First hydrogenation step of \ch{HC3N}}
\begin{figure}[h]
    \centering
    \includegraphics[width=240pt]{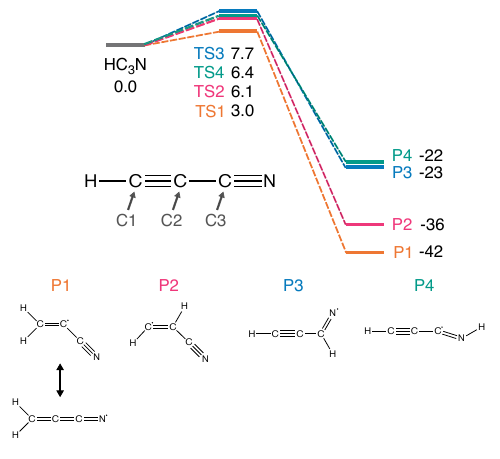}
    \caption{Energy diagram for the four possible H additions to HC$_3$N. The labeling for the transition states (TS) and products (P) corresponds to the four possible addition reactions, where 1, 2, and 3 refer to the hydrogen addition to the carbon atoms marked by C1, C2, C3 respectively and 4 refers to the addition to the nitrogen atom. The resulting structures are shown at the bottom of the figure. Electronic energies obtained at uCCSD(T)-F12/cc-pVTZ-F12//MPWB1K-D4/def2-TZVP level are given in kcal/mol and are corrected with zero-point vibrational energy obtained at MPWB1K-D4/def2-TZVP level. All energies are relative to the asymptotic state.}
    \label{fig: HC3N+H energy diagram}
\end{figure}

For the first hydrogenation step, atomic hydrogen can add to \ch{HC3N} in four different ways: to one of the three carbon atoms or to the nitrogen atom. For these four reactions the activation and reaction energies can be seen in Figure \ref{fig: HC3N+H energy diagram} and Table \ref{tab: addition energies and tc}. All four reactions are exothermic and the addition to the C1 carbon, forming the \ch{CH2CCN} (cyanoacetylenehydryl) radical, has the lowest activation energy, of 3.0 kcal/mol. 

\begin{figure}[ht]
    \centering
    \includegraphics[width=240pt]{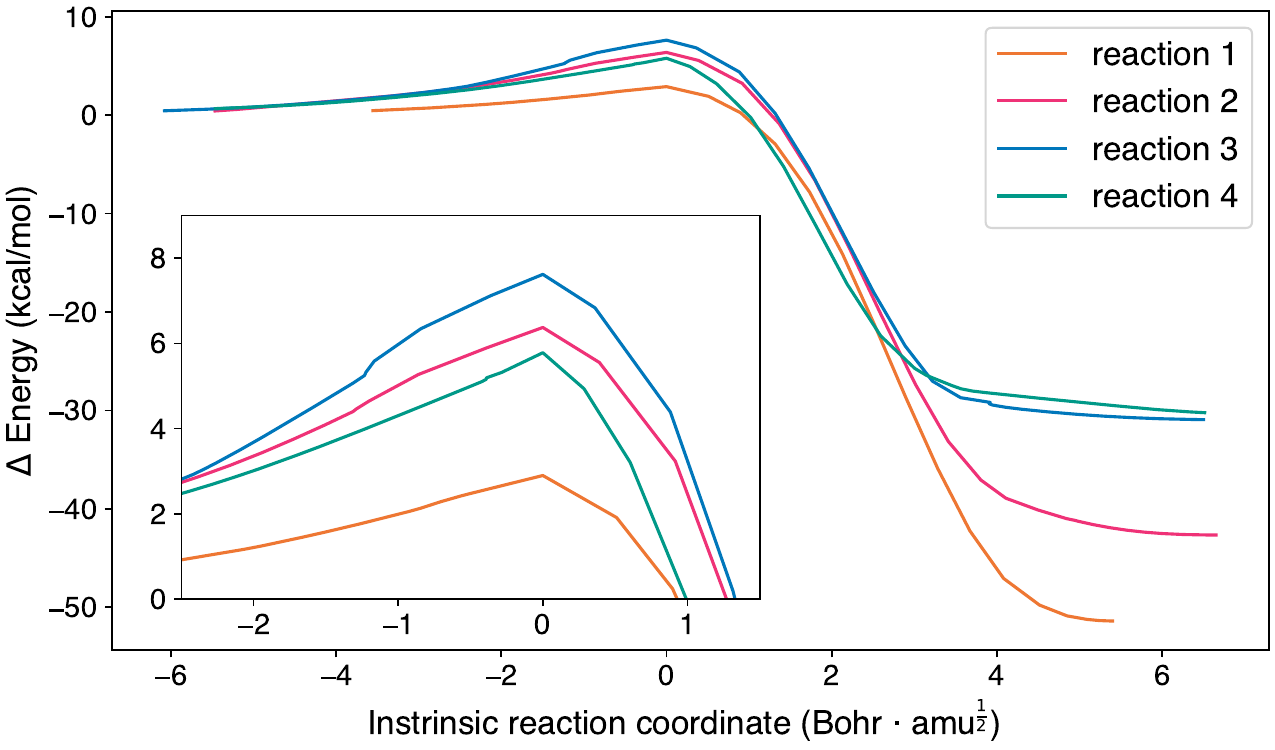}
    \caption{Energy plot obtained from IRC calculations for the \ch{HC3N} + H addition reactions 1-4. IRC calculations were performed at the MPWB1K-D4/def2-TZVP level. The energies are relative to the asymptotic state. Note that the IRC plot for reaction 1 contains fewer points as the calculation converged after 135 steps.}
    \label{fig: IRC_HC3N+H}
\end{figure}

Under cold interstellar conditions (T $\approx$ 10 K), there is very little thermal energy available to cross reaction barriers. However, reactions involving light hydrogen atoms can possibly proceed via quantum tunneling through the barrier. The temperature at which tunneling dominates the reaction rate is the crossover temperature
\begin{equation}
    T_c = \frac{\hbar\omega}{2\pi k_B},
\end{equation}\label{eq: Tc}
where $\omega$ is the absolute frequency associated with the imaginary mode of the transition state\cite{kastner_theory_2014}. The crossover temperatures for the reported reactions are shown in Table \ref{tab: addition energies and tc}. Given that the crossover temperatures for all reactions exceed the typical ISM temperature of 10 K, tunneling is very likely the dominant reaction mechanism. Another important factor influencing whether a reaction can occur via tunneling is the shape of the reaction barrier; a narrow barrier enhances tunneling. Figure \ref{fig: IRC_HC3N+H} shows that reaction 1 (the addition to the C1 carbon) has a the most narrow barrier, as well as the lowest barrier, making it the most probable reaction.

The reaction {\ch{HC3N}} + H has previously been studied by  \citet{parker_kinetics_2004}, who performed gas phase experiments to determine the reaction rates for temperatures from 200 K to 298 K, and calculations to obtain the activation energies for the four possible additions. They optimized the geometries with B3LYP/6-311G(d,p) and performed single point calculations with various levels of theory. The activation energies they report show a different trend than ours: at the CCSD(T)/6-311G(d,p)//B3LYP/6-311G(d,p) level the addition to the N atom (5.4 kcal/mol) has a slightly lower barrier than the addition to the C1 carbon (5.5 kcal/mol). This discrepancy can be partially attributed to differences in geometries, caused by the lack of a dispersion correction, and to a lesser extent caused by the use of different basis sets. However, our calculations using CCSD(T)/6-311G(d,p)//B3LYP/6-311G(d,p) result in activation energies (E$_a$=4.5 kcal/mol and E$_a$=8.2 kcal/mol for addition to the C1 and N atoms respectively) following the same trend as those in Figure {\ref{fig: HC3N+H energy diagram}}.
Besides this study, part of the potential energy surface for this reaction has gotten some attention in the context of \ch{HC3N} formation \cite{woon_rate_1997, fukuzawa_molecular_1997, huang_crossed_2000, singh_quantum_2021}. The trend in energies we found agrees with the results obtained with various other methods, for example obtained at CCSD(T)/cc-pVTZ//CISD/DZ+P level by \citet{fukuzawa_molecular_1997}. Our activation energy of 3.0 kcal/mol for the H addition to the C1 carbon is somewhat lower, but not considerably lower, than the value of 3.4 kcal/mol used in the astrochemical model used by \citet{garrod_exploring_2017}.

Besides adding to HC$_3$N, the H radical can also abstract an H atom from the \ch{HC3N} molecule. However, we found this reaction to be endothermic (by E$_r=$ 30.7 kcal/mol) and with a high barrier (E$_a=$ 31.4 kcal/mol). Thus, it is unlikely to occur in the cold ISM, in line with previous results on the hydrogen abstraction of similar species \cite{sumathi_reaction_2001}. Additionally, the KIDA database includes the reverse reaction, \ch{H2}+\ch{C3N} $\rightarrow$ \ch{H}+\ch{HC3N} \footnote{See \url{https://kida.astrochem-tools.org/reaction/5147/C3N_+_H2.html?filter=Both}} \cite{wakelam_2014_2015}. For this reaction an activation of 2.0 kcal/mol is reported, estimated from the reaction \ch{C2H}+\ch{H2}, whereas we find an activation energy of 0.6 kcal/mol. Our calculations indicate that the activation energy included in the KIDA database is too high and that our result would give a more accurate value for use in astrochemical models.

\subsection{Hydrogenation of the open shell product}
\begin{figure}[ht]
    \centering
    \includegraphics[width=180pt]{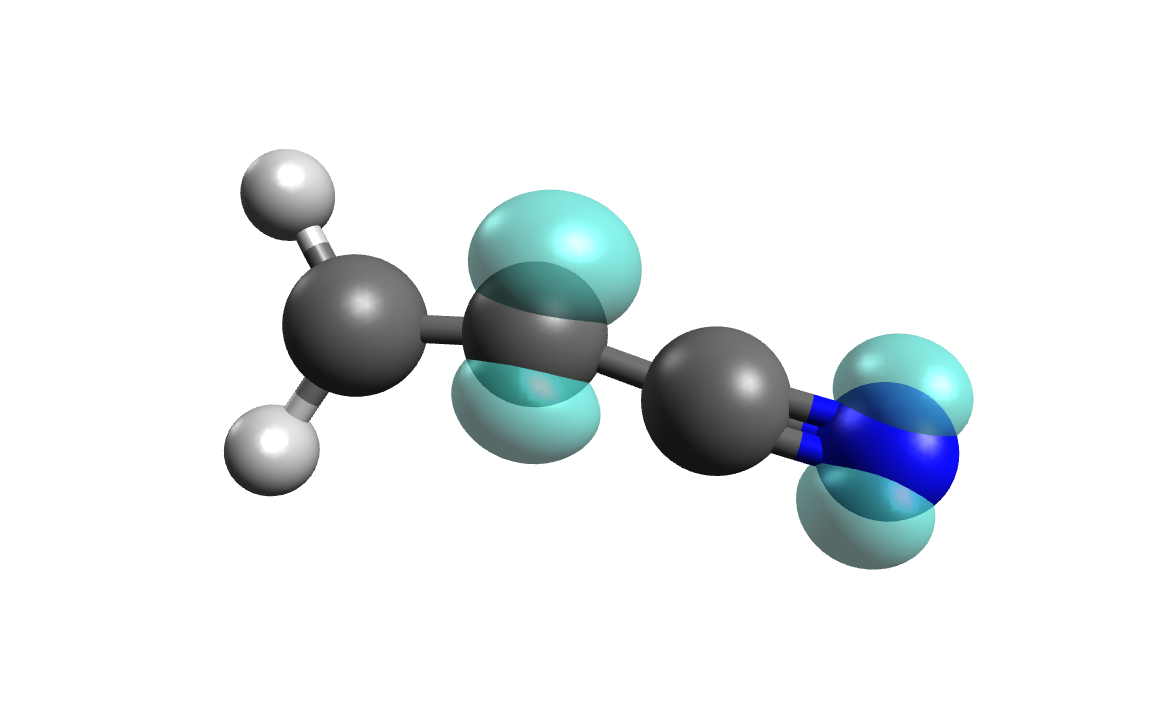}
    \caption{3D structure of \ch{CH2CCN} with spin density distribution, generated with Avogadro 1.2.0 \cite{hanwell_avogadro_2012}. The spin density is visualized using an isosurface of value 0.02 (in au).}
    \label{fig: H2C3N}
\end{figure}

Once the \ch{CH2CCN} radical is formed, it can undergo further hydrogenation. The radical's unpaired electron is partially located on the C2 carbon and partially on the nitrogen atom as indicated by its resonant Lewis structures and the distribution of the spin density on the molecule (see Figure \ref{fig: HC3N+H energy diagram} and Figure \ref{fig: H2C3N}, respectively).
Additionally, the Löwdin spin population on the C2 is 0.57 and 0.34 on the N at the MPWB1K-D4/def2-TZVP level.
The spin density plot in Figure \ref{fig: H2C3N} serves as a proxy for where a new H atom could attack via a radical-radical reaction, as shown by our broken-symmetry DFT calculations
(see Figure S4 for snapshots of one of the broken-symmetry optimizations). 
Two H-addition products can form: vinyl cyanide (\ch{CH2CHCN}), after the H-addition reaction on the C2 carbon, and \ch{CH2CCNH} (1-allenimine) through the H-N bond formation reaction.
 
Our observation that radical-radical reactions yield multiple products is in line with earlier findings \cite{lamberts_interstellar_2018, enrique-romero_quantum_2022, krim_formation_2019}. Both products found in this work, vinyl cyanide and \ch{CH2CCNH}, are subsequent starting points for a complex reaction network for further hydrogenation, as discussed in the next sections. In fact, vinyl cyanide can also be formed via pathways in the gas phase, e.g. \ch{CN}+\ch{C2H4}\cite{balucani_formation_2000}. Hence, if formed in the gas phase, vinyl cyanide could also end up freezing out on ice grains. Then, on the surface of these grains it would hydrogenate via the pathways reported in this work.

\begin{figure*}[ht]
    \centering
    \begin{subfigure}{0.49\textwidth}
        \centering
        \includegraphics[width=230pt]{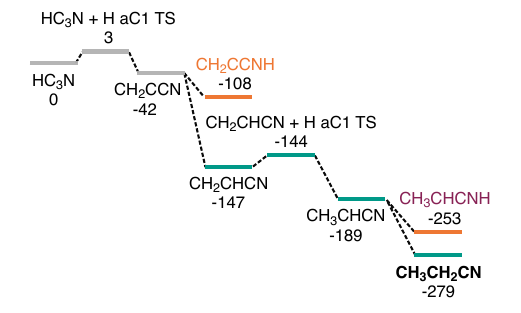}
        \caption{}\label{fig: ED_HC3N_to_EC}
    \end{subfigure}
    \begin{subfigure}{0.49\textwidth}
        \centering
        \includegraphics[width=230pt]{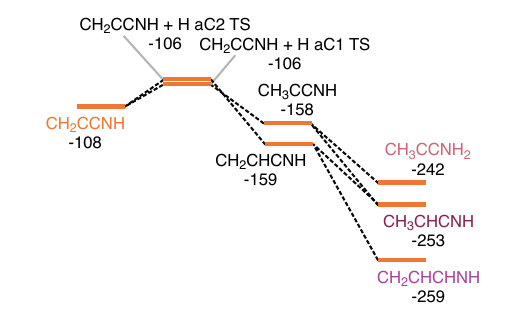}
        \caption{}\label{fig: ED_hydro_N_first_part}
    \end{subfigure}
    \begin{subfigure}{0.49\textwidth}
        \centering
        \includegraphics[width=230pt]{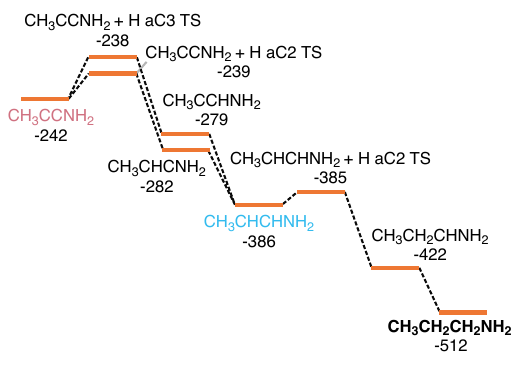}
        \caption{}\label{fig: ED_hydro_N_CH3CCNH2}
    \end{subfigure}
        \begin{subfigure}{0.49\textwidth}
        \centering
        \includegraphics[width=230pt]{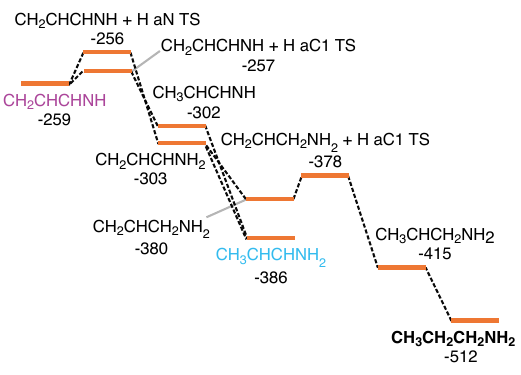}
        \caption{}\label{fig: ED_hydro_N_CH2CHCHNH}
    \end{subfigure}
    \caption{Simplified potential energy surface towards ethyl cyanide (CH$_3$CH$_2$CN, green lines, (\ref{fig: ED_HC3N_to_EC}) and propylamine (CH$_3$CH$_2$CH$_2$NH$_2$, orange lines (\ref{fig: ED_hydro_N_first_part}, \ref{fig: ED_hydro_N_CH3CCNH2}, \ref{fig: ED_hydro_N_CH2CHCHNH}). The pathways depicted here are selected based on the lowest activation energies per set of competing reactions. The species in bold are the end products of the respective pathways. The color-coded species indicate where the different pathways are connected. All energies are given relative to the initial reactant HC$_3$N in kcal/mol at the uCCSD(T)-F12/cc-pVTZ-F12//MPWB1K-D4/def2-TZVP level. The energies include ZPE and have been corrected for the total number of free hydrogen atoms.}
    \label{fig: network_ED}
\end{figure*}

\subsection{Pathway towards ethyl cyanide}

Vinyl cyanide can be further hydrogenated, see Figure \ref{fig: ED_HC3N_to_EC}. Again, the H addition to the C1 carbon (the farthest away from the N atom) is the most plausible reaction pathway since it has the lowest (2.1 kcal/mol) and most narrow barrier (see Figure S2). The resulting product, \ch{CH3CHCN} (acrylonitrilehydryl), can again react with another H atom via a radical-radical reaction, which can give two different species: ethyl cyanide (\ch{CH3CH2CN}) and \ch{CH3CHCNH} (1-propen-1-imine). Ethyl cyanide is unlikely to hydrogenate further at cold temperatures, as the addition of atomic H to the cyano-group sports high barriers of 6.5 and 9.1 kcal/mol (see Table \ref{tab: addition energies and tc}).
This agrees with earlier experimental and theoretical results on \ch{CH3CN} and experimental results on ethyl cyanide, which found that the cyano-group does not hydrogenate under cold ISM conditions \cite{nguyen_formation_2019, krim_formation_2019}.

Besides their results on ethyl cyanide, \citet{krim_formation_2019} also report experimental and theoretical results on the hydrogenation of vinyl cyanide. They found similar trends as we report here. Although their theoretical results quantitatively differ from ours, they also found the hydrogen addition to the C1 carbon to be the most likely reaction and a subsequent radical-radical reaction resulting in ethyl cyanide and \ch{CH3CHCNH}. The quantitative differences between our work and the work by \citet{krim_formation_2019} are likely caused by the different methodological approaches used: these authors ran their simulations with MP2, we did so with the functional MPWB1K-D4. In our benchmark, we have included a small study where we show that our functional provides a better picture than MP2, compared to uCCSD(T)-F12 single point references (see SI section 4).

The hydrogenation of vinyl cyanide on ice grains has been included in models by \citet{garrod_exploring_2017}, who report using activation energies of 1.2 kcal/mol and 2.6 kcal/mol for H-addition to the C1 and C2, respectively. In contrast, our calculations show activation energies of 2.2 kcal/mol and 5.1 kcal/mol for these same reactions. These differences likely affect the destruction rate of vinyl cyanide and formation rate of ethyl cyanide, and the abundances of these molecules in astrochemical models. 

\subsection{Pathways towards propylamine}

After the first hydrogenation of \ch{HC3N} into \ch{CH2CCN}, the second H-addition could yield \ch{CH2CCNH} instead of \ch{CH2CHCN}, depending on whether the hydrogen atom adds to the nitrogen or the C2 carbon. This unlocks a different branch of the reaction network shown in Figure \ref{fig: ED_hydro_N_first_part}. For the sake of clarity, we only report the most likely pathways here, selected via the lowest activation energies out of the different additions to the considered reactant. The full overview of the energetics obtained in this work can be found in Table S1.

Hydrogenation of the C1 carbon of \ch{CH2CCNH} (E$_a=1.4$ kcal/mol), forms the radical \ch{CH3CCNH} (alleniminehydryl) and the subsequent radical-radical reaction can form to possible products: \ch{CH3CCNH2} (1-propyn-1-amine) and \ch{CH3CHCNH} (1-propen-1-imine). Conversely, hydrogenation of the C2 of \ch{CH2CCNH} (E$_a=2.0$ kcal/mol) and subsequent hydrogenation of the formed radical leads to allylimine (\ch{CH2CHCHNH}) and again to \ch{CH3CHCNH}. The former has been tentatively detected in the ISM \cite{alberton_laboratory_2023}. 

From the three species discussed in the previous paragraph (\ch{CH3CCNH2}, \ch{CH3CHCNH} and \ch{CH2CHCHNH}), we have found various low barrier pathways to the fully saturated species, propylamine (\ch{CH3CH2CH2NH2}, see Figure \ref{fig: ED_hydro_N_CH3CCNH2} and Figure \ref{fig: ED_hydro_N_CH2CHCHNH}). Firstly, the hydrogenation of \ch{CH3CCNH2} (Figure \ref{fig: ED_hydro_N_CH3CCNH2}), on both the C2 carbon (E$_a=2.5$ kcal/mol) and C3 carbon (E$_a=3.3$ kcal/mol) lead to, after a radical-radical reaction, \ch{CH3CHCHNH2} (1-propen-1-amine). This species can then form propylamine, after a barrier-mediated reaction (E$_a=$ 1.1 kcal/mol) and barrierless reaction. 

Secondly, allylimine (Figure \ref{fig: ED_hydro_N_CH2CHCHNH}) has four possible hydrogen additions, of which the addition to the C1 carbon has the lowest (E$_a=1.9$ kcal/mol) and most narrow (see Figure S3) barrier. Subsequent barrierless hydrogenation leads again to \ch{CH3CHCHNH2}, which can hydrogenate to propylamine via the just discussed pathway. Additionally, the nitrogen atom of allylimine can hydrogenate via a notably low barrier of 2.9 kcal/mol. The ensuing radical-radical reaction can lead to again \ch{CH3CHCHNH2} or to \ch{CH2CHCH2NH2} (2-propen-1-amine). The C1 carbon of the latter can hydrogenate over a barrier of 2.2 kcal/mol, forming propylamine after another barrierless hydrogenation. Thirdly, \ch{CH3CHCNH}, which can both form via hydrogenation of \ch{CH3CHCN} (see Figure \ref{fig: ED_HC3N_to_EC}), hydrogenation of \ch{CH3CCNH} and \ch{CH2CHCNH} (Figure \ref{fig: ED_hydro_N_first_part}), can hydrogenate via a barrier of 3.8 kcal/mol to form the \ch{CH3CHCHNH} radical. This reaction has been omitted from Figure \ref{fig: network_ED} for clarity. The \ch{CH3CHCHNH} radical can further hydrogenate to propylamine via the pathway given in Figure \ref{fig: ED_hydro_N_CH2CHCHNH}.

\begin{table*}[ht]
\caption{Energies and crossover temperatures of the selected reactions. Reaction energies and barriers in kcal/mol at uCCSD(T)-F12/cc-pVTZ-F12//MPWB1K-D4/def2-TZVP level. ZPE corrections in parentheses  at the MPWB1K-D4/def2-TZVP. The labels aC1, aC2, aC3 refer to hydrogen addition to the C1, C2 and C3 carbon atoms respectively, aN refers to hydrogen addition to the N atom. Crossover temperatures are reported in K and are calculated using equation 1.}\label{tab: addition energies and tc}
\centering
\makebox[\textwidth][c]{\begin{tabular}{|c|c|c|c|c|} \hline
reaction name & label & \begin{tabular}[x]{@{}c@{}} reaction energy \\ (kcal/mol)\end{tabular} & \begin{tabular}[x]{@{}c@{}}barrier height\\ (kcal/mol)\end{tabular} & $T_c$ (K) \\ \hline
    \ch{HC3N} + H & aC1 & -46.9 (5.0) & 3.1 (-0.1) & 168 \\
 & aC2 & -41.4 (5.8) & 6.1 (0.0) & 218 \\
 & aC3 & -28.4 (5.3) & 7.5 (0.1) & 239 \\
 & aN & -26.8 (4.8) & 6.6 (-0.2) & 250 \\
\ch{CH2CHCN} + H & aC1 & -48.0 (5.8) & 1.5 (0.7) & 118 \\
\ch{CH3CH2CN} + H & aC3 & -27.2 (6.0) & 5.6 (0.9) & 210 \\
 & aN & -22.0 (6.3) & 8.5 (0.5) & 273 \\
\ch{CH2CCNH} + H & aC1 & -56.9 (6.3) & 0.8 (0.7) & 102 \\
 & aC2 & -58.7 (6.9) & 1.2 (0.9) & 109 \\
\ch{CH3CCNH2} + H & aC2 & -47.4 (6.7) & 1.6 (0.9) & 119 \\
 & aC3 & -43.9 (6.8) & 2.3 (1.0) & 158 \\
\ch{CH2CHCHNH} + H & aC1 & -49.2 (5.9) & 1.2 (0.6) & 115 \\
 & aN & -50.7 (6.1) & 2.1 (0.7) & 150 \\
\ch{CH3CHCNH} + H & aC3 & -56.9 (6.9) & 2.8 (1.0) & 192 \\
\ch{CH3CHCHNH2} + H & aC2 & -41.7 (6.5) & 0.2 (0.9) & 71 \\
\ch{CH2CHCH2NH2} + H & aC1 & -40.3 (5.4) & 1.5 (0.7) & 118 \\ \hline
\end{tabular}}
\end{table*}

In all of our calculations, the hydrogen addition to the unsaturated C1 carbon is the most favorable reaction, based on its lowest activation energy. Additionally, in the selection of cases we investigated, the hydrogen addition to the C1 carbon has the most narrow barrier, strengthening our finding that the H addition to the C1 carbon generally is the most likely reaction, especially considering that the reaction is tunneling-mediated. This finding agrees with results found for unsaturated molecules containing aldehyde or alcohol groups \cite{zaverkin_tunnelling_2018}. We also investigated the hydrogen abstraction from a select few closed-shell molecules in the network and these reactions are all endothermic and have high energy barriers.

Comparing the barrier heights of the different additions to the C1 carbon (see Table \ref{tab: addition energies and tc}), it can be seen that the hydrogenation of a double bond (reactions \ch{CH2CHCN} + H aC1, \ch{CH2CCNH} + H aC1, \ch{CH2CHCHCN} + H aC1, \ch{CH2CHCH2NH2} + H aC1) is easier than that of a triple bond (reaction \ch{HC3N} + H aC1). The H-addition to the C1 of \ch{HC3N}, where C1 and C2 are bonded via a triple bond, has a higher energy barrier (3.0 kcal/mol) compared to any of the additions to C1 with double bonds. This agrees with the literature on the hydrogenation of triple and double bonds \cite{zaverkin_tunnelling_2018, kobayashi_hydrogenation_2017}.

\subsection{Astrochemical implications}

This work sheds light onto the reactivity of carbon chains that are expected to be present on ice grain surfaces. We show how the unsaturated carbon atoms of cyanopolyynes can hydrogenate to saturated aliphatic chains. Additionally, we also show how the functional group, in this case a cyano-group, can hydrogenate at low temperatures. We not only found how the cyano-group of \ch{HC3N} hydrogenates via a barrierless radical-radical reaction, but also how the formed imine-group hydrogenates via similar barrierless reactions. The reported possible pathways to propylamine showcase that the imine-group most likely hydrogenates via radical-radical reactions, circumventing the high barriers (see Table S1). In these radical-radical reactions, the unpaired electron is delocalized over the molecule, with some of the electron density on the imine-group. This electron density makes the imine-group susceptible to the H addition, similar to the earlier discussed mechanism for the hydrogenation of \ch{CH2CCN} (see Figure \ref{fig: HC3N+H energy diagram} and \ref{fig: H2C3N}) and the formation of ethyl cyanide. We suggest that having multiple product channels for radical-radical reactions is possible in cases where at least one radical has a delocalized electron density. For example, preliminary calculations for the aldehyde radical \ch{H2C3OH} show that the radical electron density is distributed both around the middle C atom and the O atom (Löwdin spin population on the C2 is 0.51 and on the O 0.36 at the MPWB1K-D4/def2-TZVP level). 

This study on the reactivity of carbon chains hints at how more complex molecules might form in dark molecular clouds. The pathways we found could partially explain observations of molecules in the gas phase, such as vinyl and ethyl cyanide. These molecules could form on the ice grains and then at a later stage release to the gas phase, where they are detected \cite{fedoseev_production_2024}. 
Moreover, our findings indicate possible targets for astronomical observations. Allylimine has been tentatively detected, but other intermediates reported in this work, {\ch{CH2CCNH}}, {\ch{CH3CCNH2}}, {\ch{CH2CCHNH2}}, {\ch{CH3CHCNH}}, {\ch{CH3CHCHNH2}} and {\ch{CH2CHCH2NH2}},  could be present in the ISM.
Additionally, we propose that carbon chain hydrogenation in astrochemical models can be assumed to begin at the C1 carbon, followed by the other carbons and functional groups. Our results indicate how saturated carbon chains could form on ice grains, possible leading to precursors of prebiotic molecules, specifically fatty acids.

\section{Conclusions}\label{sec: conclusions}

In this work, we present a computational DFT study to explore the hydrogen addition reaction pathways starting from \ch{HC3N}, which lead to multiple possible products, via a complex chemical network. The possible product channels depend on where the hydrogen atom adds. These products include vinyl cyanide (\ch{CH2CHCN}), ethyl cyanide (\ch{CH3CH2CN}) and propylamine (\ch{CH3CH2CH2NH2}). The main conclusions are:

\begin{itemize}
    \item There are two main pathways: \ch{HC3N} $\rightarrow$ \ch{CH2CCN} $\rightarrow$ \ch{CH2CHCN} $\rightarrow\rightarrow$ \ch{CH3CH2CN}, and \ch{HC3N} $\rightarrow$ \ch{CH2CCN} $\rightarrow$ \ch{CH2CCNH} $\Longrightarrow$ \ch{CH3CH2CH2NH2}, where the last arrow represents many different pathways to reach the end product, propylamine.
    \item Hydrogenation of the closed shell species preferably leads to H-addition to the C1 carbon atom, that is, the C atom farthest away from the nitrogen atom, as demonstrated by  both the activation energies and barrier widths.
    \item The hydrogenation of the -CN and -CNH groups usually takes place when the molecule is in a radical state (e.g., \ch{CH2CCN}). For these open-shell molecules, the unpaired electron density is delocalized over the molecule, including the -CN or -CNH group, which makes those groups susceptible to hydrogen addition.
\end{itemize}

We suggest that these conclusions generally hold true for other carbon chains. Hydrogen addition to any carbon chain might have a preference for the carbon furthest removed from a functional group, based on both our results and literature. Likewise, different species containing functional groups otherwise stable towards hydrogen addition could still be hydrogenated via radical-radical reactions, provided there is some unpaired electron density located on the functional group.

All the reactions reported in this work have been studied as isolated systems, as a model for the actual ice grain reactions. However, the ice surface may have some effect on the reactions. This could be especially important for the radical-radical reactions, where the radical adsorption site could obstruct certain reaction pathways. Hence, future work should include \ch{H2O} and \ch{CO} surfaces and determine the effect of surface molecules on hydrogenation reactions of carbon chains. 

\begin{acknowledgement}
The authors thank the late Harold Linnartz for his impact on the field of molecular astrophysics. M.T.R. thanks Harold Linnartz for introducing him to the wondrous topic of astrochemistry. The authors thank the BSc. students Y. Versteeg and E. Goudart for their contributions to the project. J.E.R acknowledges the support of the Horizon Europe Framework Programme (HORIZON) under the Marie Skłodowska- Curie grant agreement No 101149067, “ICE-CN”. T.L. and M.T.R thank the Leiden Institute of Chemistry for the financial support, which made this research possible. Astrochemistry in The Netherlands is supported by the NWO Dutch Astrochemistry Network (grant no. ASTRO.JWST.001). 

\end{acknowledgement}

\begin{suppinfo}

The supporting information (PDF) is available free of charge and consists of:
\begin{itemize}
    \item SI section 1: Benchmark of various density functionals for the reaction \ch{HC3N} + H
    \item SI section 2: IRC plots for the reactions \ch{CH2CHCN} + H and \ch{CH2CHCHNH} + H, with accompanying explanation of how the mass-weighted coordinates are calculated.
    \item SI section 3: Snapshots of the broken-symmetry calculations for the reaction \ch{CH2CCN} + H.
    \item SI section 4: Comparative investigation into using MP2 or DFT for the TS search for the reaction \ch{CH2CHCN} + H.
    \item SI section 5: Reaction energies, barrier heights and cross-over temperatures for all reactions studied in this work.
\end{itemize}

\end{suppinfo}

\bibliography{references}

\end{document}